\documentclass[%
reprint,
superscriptaddress,
groupedaddress,
unsortedaddress,
runinaddress,
frontmatterverbose,
showpacs,preprintnumbers,
nofootinbib,
nobibnotes,
bibnotes,
amsmath,amssymb,
prstab,
prstper,
floatfix,
]{revtex4-1}
\usepackage[english]{babel}
\usepackage{hyperref}
\usepackage{xcolor}
\usepackage{graphicx}
\usepackage{dcolumn}
\usepackage{bm}
\usepackage{hyperref}

\makeatletter
\renewcommand\@make@capt@title[2]{%
  \@ifx@empty\float@link{\@firstofone}{\expandafter\href\expandafter{\float@link}}%
   {\textbf{#1}}\@caption@fignum@sep#2\quad
}%
\usepackage[normalem]{ulem}

\begin{document}

\preprint{APS/123-QED}

\title{Thermodynamic aspects of nanoscale friction}

\author{P. C. Torche}
\email{Email: pc.torche@soton.ac.uk}
\author{T. Polcar}
\author{O. Hovorka}

\affiliation{Engineering and Physical Sciences, University of Southampton, Southampton, SO17 1BJ, UK}

\date{\today}


\begin{abstract}

\noindent Developing the non-equilibrium thermodynamics of friction is required for systematic design of low friction surfaces for a broad range of technological applications. Intuitively, the thermodynamic work done by a material sliding along a surface is expected to be partially dissipated as heat and partially transformed into the change of the internal energy of the system. However, general non-equilibrium thermodynamic principles governing this separation are presently unknown. We develop a theoretical framework based on the transition state theory combined with the conventional Prandtl-Tomlinson model, allowing to set explicit expressions for evaluating the heat dissipation and internal energy change produced during the frictional stick-slip motion of a tip of a typical friction force microscope (FFM). We use the formalism to quantify the heat dissipation for a range of parameters relevant to materials in practical applications of nanoscale friction.


\end{abstract}

\maketitle



\section{\label{sec:Introduction}Introduction}

The phenomenon of mechanical friction between two materials is a paradigmatic example of out-of-equilibrium behaviour manifested by energy dissipation, memory effects, and hysteresis. At the mesoscopic level of nanoscale friction, thermal and structural fluctuations play a dominant role and typically lead to broad spectra of characteristic timescales of the underlying relaxation processes. Although the non-equilibrium thermodynamics has been considered for describing the friction and wear processes occurring at macroscopic scales \cite{Amiri2010, klamecki1984}, it remains unexplored for the fluctuation-driven nanoscale friction. Its development would be a stepping-stone towards the improved optimisation of friction losses in nano-tribological applications \cite{Krylov2008}, and is the main subject of the present study.

An example is the thermally activated stick-slip motion observed during dry friction characterised by the slow process of the mutual `stick' of two material surfaces and the fast `slip' process associated with the transient reorganisation of their relative surface atomic configurations \cite{Krylov2008}. Stick-slip mechanism of friction has been studied by friction force microscopy (FFM), which allowed quantifying its dependence on the temperature. This led to the discovery of thermolubricity at high temperatures \cite{Barel2010, Barel2011, Schirmeisen2006, Polcar2014}, understanding the effect of humidity and oxidation in metals \cite{Polcar2005}, or revealing positive \cite{Riedo2004, Riedo2003, Sills2003, Prioli2003, Gnecco2000, Jansen2010, Bennewitz2001, Gourdon1997, Barel2010, Barel2011} or negative \cite{Barel2010, Barel2011} logarithmic velocity dependence of friction in a selected temperature range.
A thermodynamic description of the nanoscale stick-slip motion has been considered recently using the Langevin dynamics-based description of friction \cite{wang_energy_2015,pellegrini_thermally_2019}. The principle inherently relies on the fact that information about the heat and work is naturally contained in the fluctuating trajectory followed by the FFM tip during its sliding motion along the material surface, and could be extracted from it through appropriate thermodynamically consistent mathematical framework applicable to fluctuating systems. Such a general mathematical framework, which allows to systematically quantify the irreversible processes occurring in nanoscale systems driven by thermal fluctuations, has been developed recently and is referred to as modern stochastic thermodynamics \cite{Seifert2012,VanDenBroeck2015,Tome2012}. 

In this article we build on the earlier work \cite{wang_energy_2015,pellegrini_thermally_2019} from the perspective of the stochastic thermodynamics, and derive explicit expressions for entropy, which directly allow evaluating the heat produced in the system and heat transferred into the environment during the stick-slip friction process. We show that these expressions are consistent with the first thermodynamic law relating the thermodynamic work and internal system energy. Our analysis is based on the one-dimensional Prandlt-Tomlinson model as the simplest model of a single asperity friction and includes thermal activation through the Kramers transition state theory \cite{Prandtl1928,Tomlinson1929,Krylov2008,Gnecco2001}. This leads to the description of the thermally activated friction process via a system of coupled ordinary differential equations for state (asperity) probabilities, which allows applying the concepts of the stochastic thermodynamics to evaluate the individual thermodynamic variables. The formalism is then applied to quantify the heat produced during friction processes at different temperatures, sliding velocities, and energy potentials.


\begin{figure}[t]{\columnwidth 0pt}
\includegraphics[width=\columnwidth]{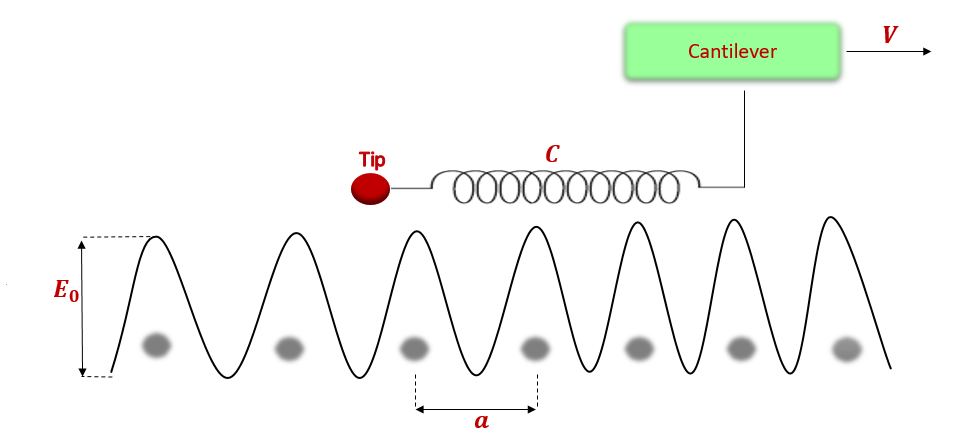} 
\caption{A sketch of the principle of the 1D  Prandtl-Tomlinson model of FMM experiment. Typical values of the parameters are $E_0\sim 0.1-1$ eV, $V$ =1 nm/s$-$1 m/s, $C=0.1-50$ N/m, and $a\approx 0.3$ nm \cite{Dong2013}.} \label{fig:fig1a}
\end{figure}

\section{\label{sec:model}Model}

\noindent We first overview the formalism and introduce the terminology essential for the development of the thermodynamic formalism in subsequent sections. The one-dimensional Prandtl-Tomlinson model describes a tip sliding in one direction on a static substrate as illustrated in FIG. \ref{fig:fig1a}. The model is defined by the following potential energy function:
\begin{equation}
\label{eq:energy}
u(x,t) = -E_0\cos\left(\frac{2\pi}{a} x(t)\right)+\frac{C}{2}(x(t)-Vt)^2
\end{equation}
The variable $x=x(t)$ describes the position of a tip at the time $t$. The first term in Eq. (\ref{eq:energy}) corresponds to the energy of substrate-tip interaction, where $E_0$ is the interaction strength and $a$ the spacing between the surface atoms. The second term corresponds to the elastic energy of the cantilever dragging the tip at a constant velocity $V$, where $C$ is the elastic energy constant in the $x$-direction.
The total force acting on the tip at the position $x$ is $f_\mathrm{tot}(x, t) = -\partial u/\partial x = f_s(x) + f(x, t)$, where:
\begin{subequations}
\begin{equation}\label{eq:forces}
f_s(x)=-\frac{2\pi}{a}E_0\sin\left(\frac{2\pi}{a}x\right)
\end{equation}
\begin{equation}\label{eq:force}
f(x, t) = - C(x-Vt)
\end{equation}
\end{subequations}
Eq. (\ref{eq:forces}) defines the force due to the substrate-tip interaction and Eq. (\ref{eq:force}) the elastic force exerted on the tip by the cantilever, respectively.

Stable states $m$ of the system are defined as the tip positions $x_m(t)$ for which the total restoring force is zero, i.e. $f_\mathrm{tot}=0$, and correspond to the local energy minima in Eq. (\ref{eq:energy}) defining the state energies denoted as $u_m(t) = u(x_m, t)$. These states identify the positions accommodating the tip in the absence of external driving and thermal fluctuations. Global energy minimisation of Eq. (\ref{eq:energy}) at a given time $t$ allows to identify all available states $m=1,\dots, M$. The total number of states $M$ may vary in time, depending on the values of $E_0$, $a$, and $C$.

\emph{Thermally activated dynamics.}
%
For the purpose of this study, we adopt the rate theory framework typically used in the description of the over-damped dynamics of the stick-slip motion \cite{Krylov2005} where thermal activation is viewed as a Markovian random hopping process over energy barriers separating the different states $m$. The rates of thermally activated transitions between the states are dependent on the energy barriers separating them. If $\Delta u_{mn}$ denotes the energy barrier separating two neighbouring states $m$ and $n$, the rate of the transition from $n$ to $m$ follows the Arrhenius law:

\begin{equation}
\label{eq:transition-rates}
\omega_{mn}(t)=f_0 \exp \Big(-\frac{\Delta u_{mn}(t)}{k_BT} \Big)
\end{equation}
where $f_0$ is the attempt frequency setting the characteristic timescale of thermal relaxation processes, $T$ is the temperature, and $k_B$ the Boltzmann constant.
The time evolution of the system is then studied by evaluating the probabilities of states $m$ at different time $t$ by solving the so-called Master equation:
\begin{equation}
\label{eq:master-equation}
\dfrac{dp_m(t)}{dt}=\sum\limits_{n}\Big( \omega_{mn}(t)p_{n}(t)-\omega_{nm}(t)p_m(t) \Big)
\end{equation}

In this work we consider only the transitions occurring between the immediately neighbouring states (asperities), i.e. the transitions from $n$ to $m$ where $m = n-1$ or $n+1$. The rate of these transitions $\omega_{mn}$ is determined by the energy barrier $\Delta u_{mn} = u_{nm}^\mathrm{max}-u_n$, where $u_{nm}^\mathrm{max}$ corresponds to the energy maximum located between the states $m$ and $n$. The transition rates $\omega_{mn}$ are zero for more distant states. This assumption is justifiable for over-damped systems when the transitions between distant states are expected to be rare. However, the formalism is fundamentally not restricted to the nearest neighbor state transitions, and longer range transitions can naturally be included in Eqs. (\ref{eq:transition-rates})-(\ref{eq:master-equation}) as well. Given that the focus of the present study is on the development of appropriate thermodynamic formalism to describe thermally activated friction processes, we postpone the question of the effect of such longer range transitions to future work.

\begin{figure}[t]{\columnwidth 0pt}
\includegraphics[width=\columnwidth]{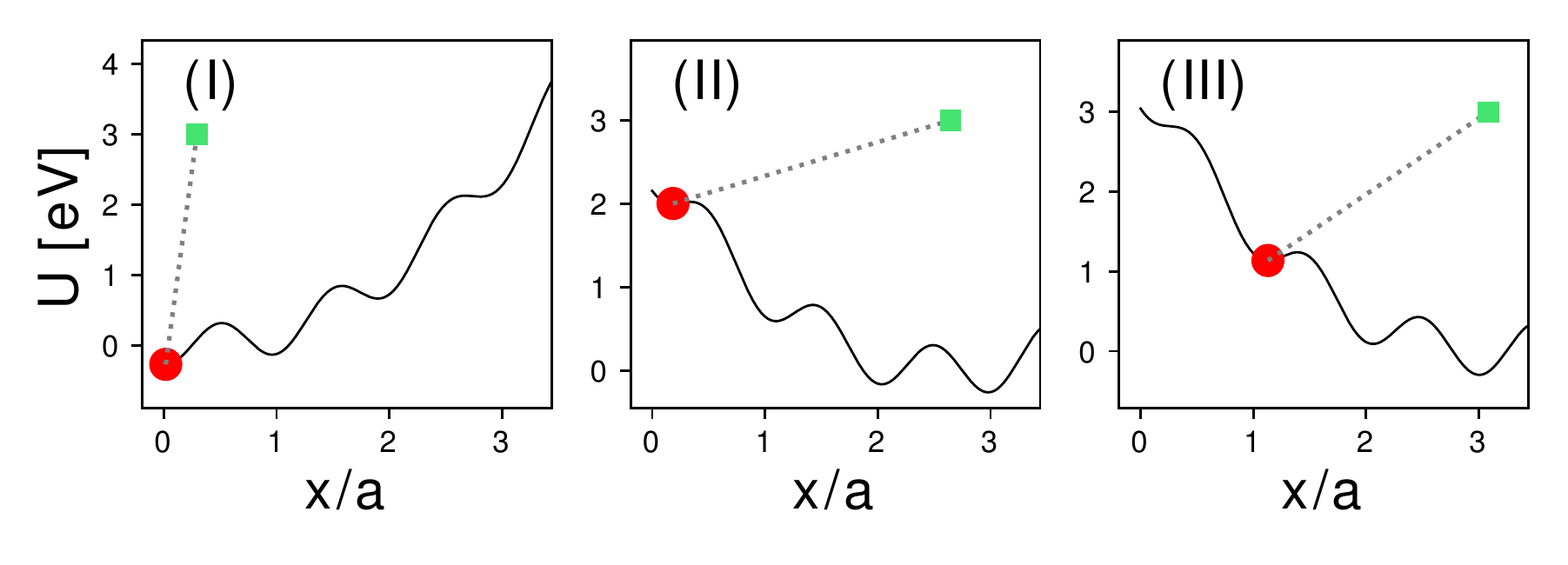} 
\caption{Time-dependent energy in the Prandtl-Tomlinson model obtained from Eq. (\ref{eq:energy}) for $\gamma=E_0(2\pi)^2/Ca^2 = 16.7$, at three different times $t_1$ (I), $t_2$ (II), $t_3$ (III), such that $t_1<t_2<t_3$. The times between (I) and (II) correspond to the system being in the `stick' state, and at some point between (II) and (III) the `slip' occurs. The red circle corresponds to the position of the tip, and the green square corresponds to the position of the cantilever at a given time instant.} \label{fig:fig1b}
\end{figure}

\emph{Algorithm.}
To set up Eqs. (\ref{eq:transition-rates}) and (\ref{eq:master-equation}) it is necessary to determine all stable states $m$ and energy barriers $\Delta u_{mn}$ by identifying all energy minima and maxima available at any given time $t$. Inspecting Eq. (\ref{eq:energy}) shows that more than one energy minimum exists in the stick-slip motion parameter range when $\gamma = E_0(2\pi)^2/Ca^2 > 1$. Moreover, due to the explicit time dependence in the second term in Eq. (\ref{eq:energy}) the energy minima and maxima continually evolve in time during the progression of the sliding motion, as illustrated in FIG. \ref{fig:fig1b}, may disappear and reappear, and their total number is not conserved.
To capture the variable nature of the energy landscape we developed the following approach. Assuming the time duration of a hypothetical FFM experiment extends from $0$ to $t_\mathrm{max}$, we first divide this interval into time instants $t_k$, where $k=0,\dots, N$ such that $t_0=0$, $t_N=t_\mathrm{max}$, and $t_k-t_{k-1}=\Delta t$ with $\Delta t$ being a small time-step. We then set a global time counter to $t_0$ and evaluate and store all states $x_m(t_0)$ obtained by extensively minimising Eq. (\ref{eq:energy}) for $t=t_0$. This procedure is repeated for every subsequent time $t_k\le t_N$, and the identified states $x_m(t_k)$ are stored as arrays labeled by the time index variable $k$. Thus this procedure allows pre-computing a time-ordered arrays of states spanning the entire time interval from 0 to $t_\mathrm{max}$, from which all the states corresponding to a given time instant can be directly accessed through the index $k$.

\emph{Deterministic solution}. Once all the states have been pre-computed, Eq. \eqref{eq:master-equation} can be set up by applying a mask setting all transition rates corresponding to transitions between the states available at a given time $t_k$ to $\omega_{mn}$, and to zero otherwise. The mask and the values of the $\omega_{mn}$ are updated at every time step $k$, and the solution of the system of ordinary differential equations in Eq. (\ref{eq:master-equation}) starting from the initial condition $p_m(t_0=0)$ is obtained by numerical integration using LSODA/Runge-Kutta algorithm \cite{seiler_numerical_1989}.

\emph{Stochastic solution.} Rather than evaluating the probability distribution, Eq. \eqref{eq:master-equation} can be solved by generating individual randomized trajectories of state variables by using kinetic Monte-Carlo methods. Here we set up the so-called fixed time-step kinetic Monte-Carlo method \cite{Jansen2012}, which only requires the knowledge of the possible transition paths and the energy barriers quantifying the associated transition rates through Eq. (\ref{eq:transition-rates}), as they were determined above. The actual thermally fluctuating trajectories followed by the FFM tip consistent with Eqs. (\ref{eq:transition-rates})-\eqref{eq:master-equation} can then be evaluated directly. 

Deterministic and stochastic solutions will be compared below assuming the following realistic parameter values: $E_0=0.3$ eV, $a=0.25$ nm, $f_0=19.5$ kHz as determined experimentally \cite{Riedo2003}, and $k_B=8.617\times 10^5$ eV/K. The default velocity $V=10$ nm/s, substrate corrugation $E_0=0.3$ eV, and temperature $T=300$ K were varied in the parameter sweep in the intervals 0.1$-$100 nm/s, 0.2$-$0.8 eV, 200$-$500 K, respectively, which corresponds to dimensionless $\gamma=E_0(2\pi)^2/Ca^2=11.2-45$. All calculations below assume the initial condition $x=0$ at $t=0$.
 
%
\begin{figure}[t!]
\includegraphics[width=\columnwidth]{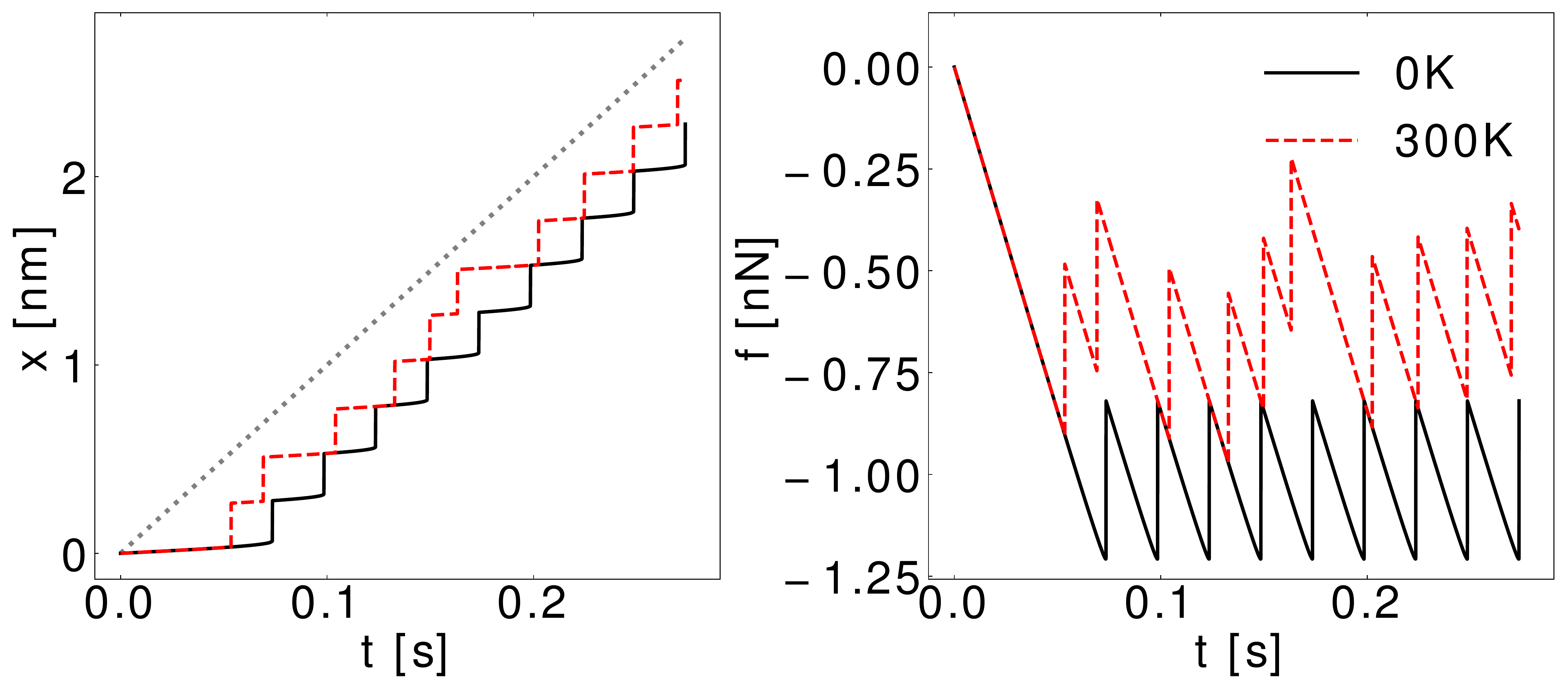}
  \caption{
  The position of the tip (left) and lateral force (right) as a function of time at 0 K and 300 K. The position of the cantilever is highlighted by a grey dashed line in the left figure. These trajectories were produced by a kinetic Monte-Carlo algorithm consistent with Eqs. (\ref{eq:transition-rates})-(\ref{eq:master-equation}). 
  }
 \label{fig:x_f_traject}
 \end{figure}
\section{\label{sec:force} Friction force}

\noindent 
Fig. \ref{fig:x_f_traject} shows the simulated instantaneous trajectories of the tip position $x_m(t)$ and the force $f_m(t)$ computed by using the kinetic Monte-Carlo method.
The solid lines in FIG. \ref{fig:x_f_traject} relate to a zero temperature case when thermal fluctuations are absent and the sliding motion is fully deterministic, as expected. 
%
At non-zero temperature, when thermal fluctuations play a role, the state transitions gain non-zero probability even for finite energy barriers. Then, statistically, the stick-slip transitions occur earlier in comparison to the zero-temperature case, which leads to the shift of the tip trajectory to the shorter timescale range as shown in FIG. \ref{fig:x_f_traject}(a), and to the reduction of the magnitude of the lateral force, i.e. reduced friction, as shown in FIG. \ref{fig:x_f_traject}(b).

The mean tip position and the mean lateral friction force can be obtained by averaging over a large number of randomised trajectories or as expectation values over the state probabilities $p_m(t)$ obtained by solving Eq. (\ref{eq:master-equation}): 
\begin{equation}
\label{eq:expectation-X}
X(t) = \sum_m p_m(t) x_m(t)
\end{equation}
and:
\begin{equation}
\label{eq:expectation-f}
F(t) = -\sum_m p_m(t) f_m(t)
\end{equation}
where according to Eq. (\ref{eq:force}) the $f_m(t)=f(x_m,t)$ is the force exerted by the substrate on the tip, as reflected by the minus sign, if the system is in the state $x_m(t)$. The friction force can ultimately be defined as an average of $F$ over the measurement time $t_\mathrm{max}$:
\begin{equation}
\label{eq:frictionf}
    \bar F = \frac{1}{t_\mathrm{max}}\int_0^{t_\mathrm{max}} F(t)dt
\end{equation}
as is conventional \cite{Krylov2008}.

\begin{figure}
\includegraphics[width=\columnwidth]{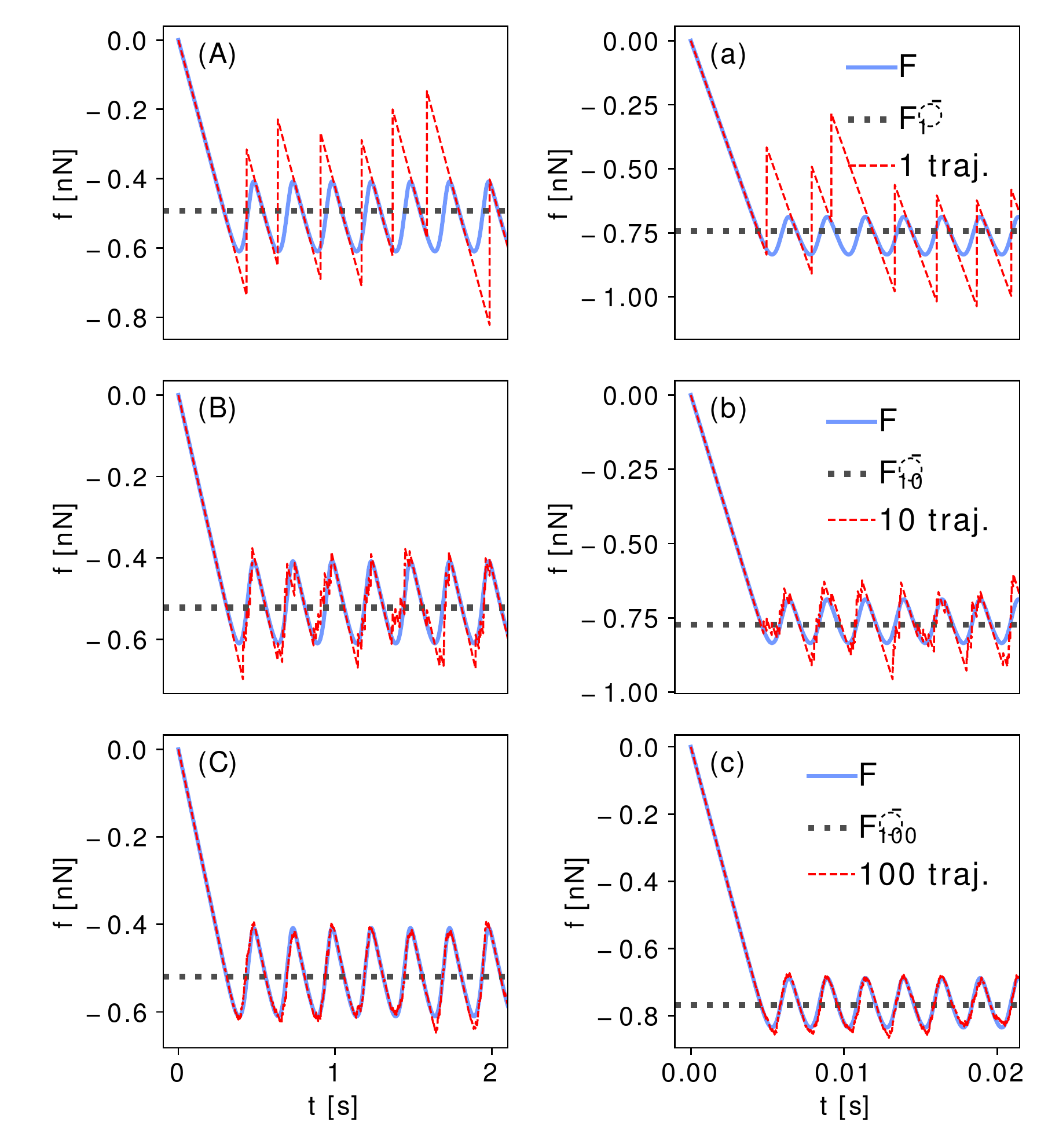}
  \caption{Average force (Eq. \ref{eq:force}) at 300 K produced over a variable number of the individually generated stochastic trajectories for velocities 1 nm/s (left) and 100 nm/s (right).  Blue continuous line: expectation value of the force, computed as Eq. (\ref{eq:expectation-f}). Red dashed line: (A)-(a) Force evaluated over a single trajectory, (B)-(b) averaged over 10 independent trajectories of a typical FFM experiment, and (C)-(c) averaged over 100 independent trajectories and closely resembling the solution obtain from Eq. (\ref{eq:expectation-f}) (blue continuous line). All trajectories were produced by a kinetic Monte-Carlo algorithm consistent with Eqs. (\ref{eq:transition-rates})-(\ref{eq:master-equation}). Dotted grey line is the time average force (friction) of the red dashed trajectories.
  }
\label{fig:f_trajectories}
\end{figure}

FIG. \ref{fig:f_trajectories} illustrates the comparison between the trajectory averaging and the direct solution based on Eq. (\ref{eq:expectation-f}) assuming slow and fast dragging speeds $V$. Namely, figures (b)-(B) and (c)-(C) show, respectively, averages over 10 and 100 of such stochastic trajectories and confirm that sufficient averaging recovers the solutions consistent with Eq. (\ref{eq:master-equation}) and Eq. (\ref{eq:expectation-f}). In addition, FIG. \ref{fig:f_trajectories} confirms the observations from previous studies \cite{Jansen2010} that the magnitude of the mean friction force increases with the increasing $V$.


%
%

\section{\label{sec:energy-change} Thermodynamic considerations}

\noindent 
FIGs. \ref{fig:x_f_traject} and \ref{fig:f_trajectories} show that the system initially follows a transient behaviour observed in the early stages of the friction process, which gradually settles into a repetitive steady state. Any consistent non-equilibrium thermodynamic framework needs to allow quantifying the thermodynamic work and heat produced during both the transient and steady state stages of the friction process.

\subsection{First law of thermodynamics}

\noindent Fundamentally, from the point of view of statistical mechanics, the Master equation (\ref{eq:master-equation}) inherently implies the assumption of a small fluctuating system connected to an infinite heat bath of temperature $T$ allowing for the heat transfer, consistent in equilibrium with the standard Boltzmann distribution. In order to develop the thermodynamic description of the friction process, Eq. (\ref{eq:master-equation}) needs to be combined with the first thermodynamic law:
\begin{equation}
\label{eq:first_law}
	\dfrac{dU}{dt} = \dfrac{ \delta W}{\delta t} - \dfrac{\delta Q}{\delta t}
\end{equation}
which relates the change of the internal energy of the system ($U$), thermodynamic work ($W$), and the heat exchanged with the heat bath ($Q$).

The internal energy of the system can be defined similarly to Eqs. (\ref{eq:expectation-X}) and (\ref{eq:expectation-f}) as the mean value of state energies $u_m(t)$:
\begin{equation}
\label{eq:expectation-U}
U(t) = \sum_m p_m(t) u_m(t).
\end{equation}
Differentiating this expression with respect to time gives:
\begin{equation}
\label{eq:derivative-energy}
	\dfrac{dU}{dt}= \sum_m p_m  \dfrac{du_m}{dt} + \sum_m  \dfrac{dp_m}{dt} u_m.
\end{equation}
In the first sum, the time derivative can be expressed by using the chain rule as $du_m/dt = (\partial u_m/\partial x)dx/dt + \partial u_m/\partial t$. The first partial derivative is zero because the stable states $x_m$ correspond to energy minima. The second partial derivative can be expressed in view of Eqs. (\ref{eq:energy}) and (\ref{eq:force}) as $\partial u_m/\partial t = -C(x_m-Vt)V = f_m(t)V$, which upon inserting in the first term in Eq. (\ref{eq:derivative-energy}), averaging by using Eq. (\ref{eq:expectation-f}), and arranging leads to: 
\begin{equation}
\label{eq:work}
\sum_m p_m \dfrac{du_m}{dt} = -FV \equiv \dfrac{ \delta W}{\delta t}
\end{equation}
Eq. (\ref{eq:work}) postulates the mean thermodynamic work per unit time consistently with the standard notions that the incremental mechanical work is the product of velocity and force. It relates the work to the various parameters and time-dependent state probabilities entering in Eqs. (\ref{eq:transition-rates}) and (\ref{eq:master-equation}), and allows its quantification during both the transient and steady state stages of the friction process.

The interpretation of Eq. (\ref{eq:work}) as mean work is in fact quite intuitive, given that the term on the left-hand side is nothing but the expectation value of the power input delivered from the cantilever. However, justifying its thermodynamic meaning as work requires confirming the energy conservation in Eq. (\ref{eq:first_law}). Then, given that the first term in Eq. (\ref{eq:derivative-energy}) is suggested to act as thermodynamic work, it is necessary to show that the second term in Eq. (\ref{eq:derivative-energy}) can be interpreted as the heat flow between the system and its surroundings.

The second sum in Eq. (\ref{eq:derivative-energy}) can be expressed through Eq. (\ref{eq:master-equation}) inserted in place of the time derivative of probability. Using the relation $u_{m}-u_{n} = k_BT \ln(\omega_{nm}/\omega_{mn})$, which follows directly from Eq. (\ref{eq:transition-rates}), and arranging gives:
\begin{equation}
\label{eq:second_term}
\begin{split}
    \sum_m  \dfrac{dp_m}{dt} u_m =
    k_BT\sum_{n>m} (\omega_{mn}p_n-\omega_{nm}p_{m}) \ln \frac{\omega_{mn}}{\omega_{nm}}
\end{split}
\end{equation}
As we show below, this expression can indeed be related to the entropy flow between the system and the environment, i.e. to the heat transferred to the environment.

 \subsection{Entropy production: heat generation}

\noindent Non-equilibrium entropy $S(t)$ is defined by the standard Gibbs formula \cite{Tome2012}:
\begin{equation}
\label{eq:S}
S  = -k_B \sum_m p_m(t) \ln p_m(t)
\end{equation}
where $p_m(t)$ are the state probabilities. The time derivative of $S$ is:
\begin{equation}
\label{eq:dS}
 \dfrac{d S }{dt} = k_B \sum_{m} (\omega_{mn}p_{n}-\omega_{nm}p_{m}) \ln \dfrac{p_{n}}{p_m}
\end{equation}
which results from using Eq. \eqref{eq:master-equation} during the differentiation of Eq. (\ref{eq:S}) and subsequent algebraic manipulations. It is conventional in non-equilibrium thermodynamics to split the total entropy change into the so-called entropy production $\delta_iS/dt$ and entropy flow $\delta_eS/dt$ \cite{prigogine1967}:
\begin{equation}
\label{eq:relation_three_entropies}
  \dfrac{dS }{dt} = \dfrac{\delta_iS}{\delta t} - \dfrac{\delta_eS}{\delta t}
\end{equation}
The special notation using the $\delta$ symbol is to emphasise that the entropy production and flow are not state variables and depend on the path connecting the starting and the end states associated with the system evolution, while the total entropy change is a state variable and path-independent.

\emph{Entropy production.} Entropy production quantifies the extent of irreversible processes occurring within the system. The second law of thermodynamics states that $\delta_iS/\delta t \ge 0$, where the equality holds only of reversible (equilibrium) processes. The expression for entropy production for systems described by Master equations has been postulated earlier \cite{Schnakenberg1976}:
\begin{equation}
\label{eq:diS}
   \dfrac{\delta_iS}{\delta t}=
   k_B \sum_{n>m}(\omega_{mn}p_n-\omega_{nm}p_{m}) \ln \dfrac{p_n\omega_{mn}}{p_m\omega_{nm}}
\end{equation}
It is straightforward to see that this expression satisfies the second law of thermodynamics, since the signs of the term within the parentheses and the logarithm always balance out leading to a positive product of both terms in the sum.

\emph{Entropy flow.} To obtain the expression for the entropy flow we use Eq. (\ref{eq:relation_three_entropies}) and subtract Eq. (\ref{eq:diS}) from (\ref{eq:dS}), which after arranging gives:
\begin{equation}
\label{eq:entropy_flow}
\begin{split}
   \dfrac{\delta_eS}{\delta t} =
    k_B\sum_{n>m} (\omega_{mn}p_n-\omega_{nm}p_{m}) \ln \frac{\omega_{mn}}{\omega_{nm}}
\end{split}
\end{equation}
This result reproduces the expression obtained independently in Eq. (\ref{eq:second_term}) except for the missing temperature pre-factor. For closed systems considered here, able to exchange only heat with the environment, the entropy flow can be related to heat through the Carnot-Clausius theorem stating that \cite{degroot1963}:
\begin{equation}
\label{eq:heat}
\dfrac{ \delta_eS}{\delta t}\equiv \frac{1}{T}\frac{\delta Q}{\delta t}
\end{equation}
where $\delta Q$ is the heat exchanged between the system and the environment. This recovers the missing $T$ factor in Eq. (\ref{eq:second_term}) and adds to it the meaning of the heat flow between the system and the surroundings, i.e.:
\begin{equation}
\label{eq:heat2}
\begin{split}
    \sum_m  \dfrac{dp_m}{dt} u_m = T\dfrac{\delta_eS}{\delta t} = \dfrac{\delta Q}{\delta t} 
\end{split}
\end{equation}
which completes the definition of the first law of thermodynamics stated in Eq. ({\ref{eq:first_law}}) and allows interpreting the thermodynamic work and heat.

Generally speaking, entropy flow given by Eqs. (\ref{eq:entropy_flow}) and (\ref{eq:heat}) is the entropy contribution supplied to the system by the heat bath. It can be positive or negative depending on the interaction of the system with its surroundings. The sign convention used here is positive when the flow is from the system into the surroundings.
It is also useful, following Eq. \eqref{eq:heat}, to define the expression:
\begin{equation}
\label{eq:Qp}
  \dfrac{ \delta Q_p}{\delta t} = T  \dfrac{\delta_i S}{\delta t}
\end{equation}
which relates the entropy production to the heat produced by the system itself during the irreversible internal processes.

\begin{figure}[t]
\centering
\includegraphics[width=0.7\columnwidth]{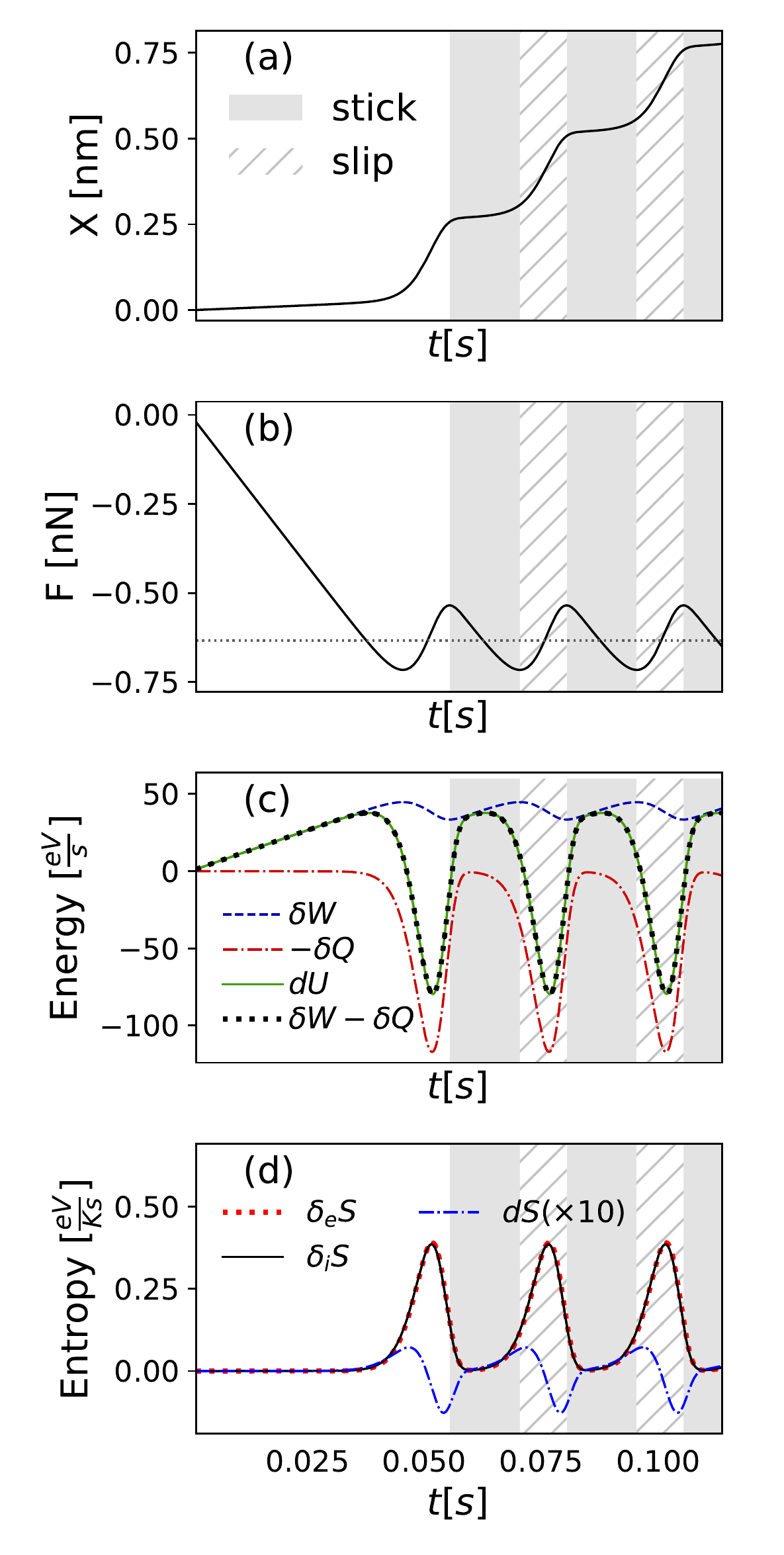}
\caption{Expectation values of: (a) position of the tip, (b) lateral force of the substrate, (c) heat exchanged with the surroundings, energy change, work, (d) entropy production, entropy flow, and entropy change of the tip, at 300 K and $V=10$ nm/s. In `stick' stages: (a) the position remains almost constant, (b) the force increases in magnitude, (c) work increases, small amounts of heat is dissipated, (d) entropy of the system increases. In `slip' stages: (a) the position increases about one lattice spacing, (b) the force relaxes, (c) work is reduced, heat dissipation has its peak, (d) entropy of the system decreases to it minimum, and entropy production and entropy flow have a peak. The total entropy change has been magnified by a factor of 10 for better visualisation. 
}
\label{fig:example_300K}
\end{figure}


\section{\label{sec:results}Results and Discussion}

\noindent We now apply the developed thermodynamic formalism to study various thermodynamics aspects of the nanoscale friction based on specific systems.

FIG. \ref{fig:example_300K}(a)-(b) shows the mean position of the tip and the lateral force assuming temperature $T = 300$ K and dragging velocity $V=10$ nm/s. The transient behaviour in the early stages of the time evolution turns into the steady state where the `stick' and `slip' stages alternate periodically.
FIG. \ref{fig:example_300K}(c) shows the corresponding time evolution of the incremental thermodynamic work and heat flow between the system and heat bath. As suggested by the dotted and solid lines, the sum of the work and heat equals to the internal energy change during both the transient and steady state stages of the sliding motion, as expected based the first thermodynamic law Eq. (\ref{eq:first_law}).
FIG. \ref{fig:example_300K}(d) illustrates that the entropy production is always non-negative as expected based on the second thermodynamic law. The contributions to the change of the entropy terms in Eq. (\ref{eq:relation_three_entropies}) are zero during the initial transient period and increase rapidly during `slip' events dominated irreversibility. Although the total entropy change is rather small on the scale of the entropy production and entropy flow, it remains non-zero during the steady state stage of the friction process. 

\begin{figure}
\centering
\includegraphics[width=\columnwidth]{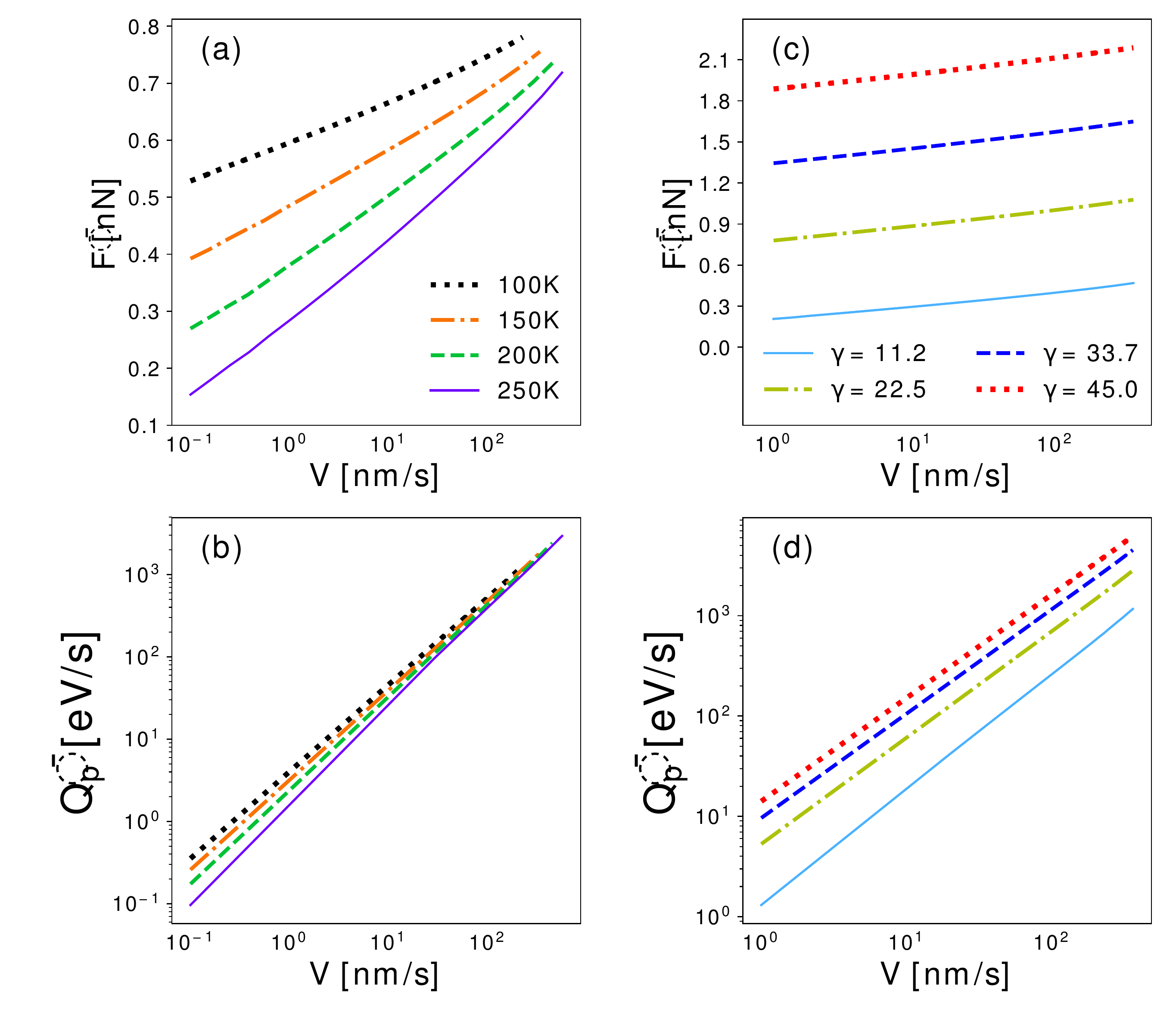}
  \caption{$\bar F$-$V$ and $\bar Q_p$-$V$ relations. Friction-velocity relation is logarithmic while heat-velocity behaviour is linear. $\bar F$ and $Q_p$ are calculated for temperatures $200K-500K$, substrate corrugation $E_0\approx 0.2eV-0.8eV$ ($\gamma$ between $11.2-45$) and velocities $0.1nm/s \leq V \leq V_{max}$ \cite{SI}. The rest of model parameters are fixed at the default values. }
\label{fig:fig5}
\end{figure}

FIG. \ref{fig:fig5}(a) shows the mean friction force calculated by using Eq. (\ref{eq:frictionf}) as a function of the magnitude of the dragging velocity for different temperatures. The friction force increases with the increasing velocity and decreasing temperature, due to the reduction of the effect of thermal fluctuations. Similarly, as shown in FIG. \ref{fig:fig5}(c), the friction force is increasing with increasing substrate corrugation due to the relative enhancement of the substrate-tip interaction. FIG. \ref{fig:fig5}(b) and (d) show similar dependence for the average heat produced in the system computed as $\bar Q_p = (1/\Delta t) \int_t^{t+\Delta t} \delta_p Q/\delta t\,dt$, where the time interval $(t, t+\Delta t)$ is chosen to include a sufficient number of stick-slip events for averaging. Thus the average heat produced during the stick-slip events increases with the velocity and with the substrate corrugation and decreases with the increasing temperature. In addition, while the velocity dependence of the friction force follows logarithmic trend, the variation of the heat appears to be linear over many decades of the dragging velocities.

\begin{figure}[t!]
\centering
\includegraphics[width=0.8\columnwidth]{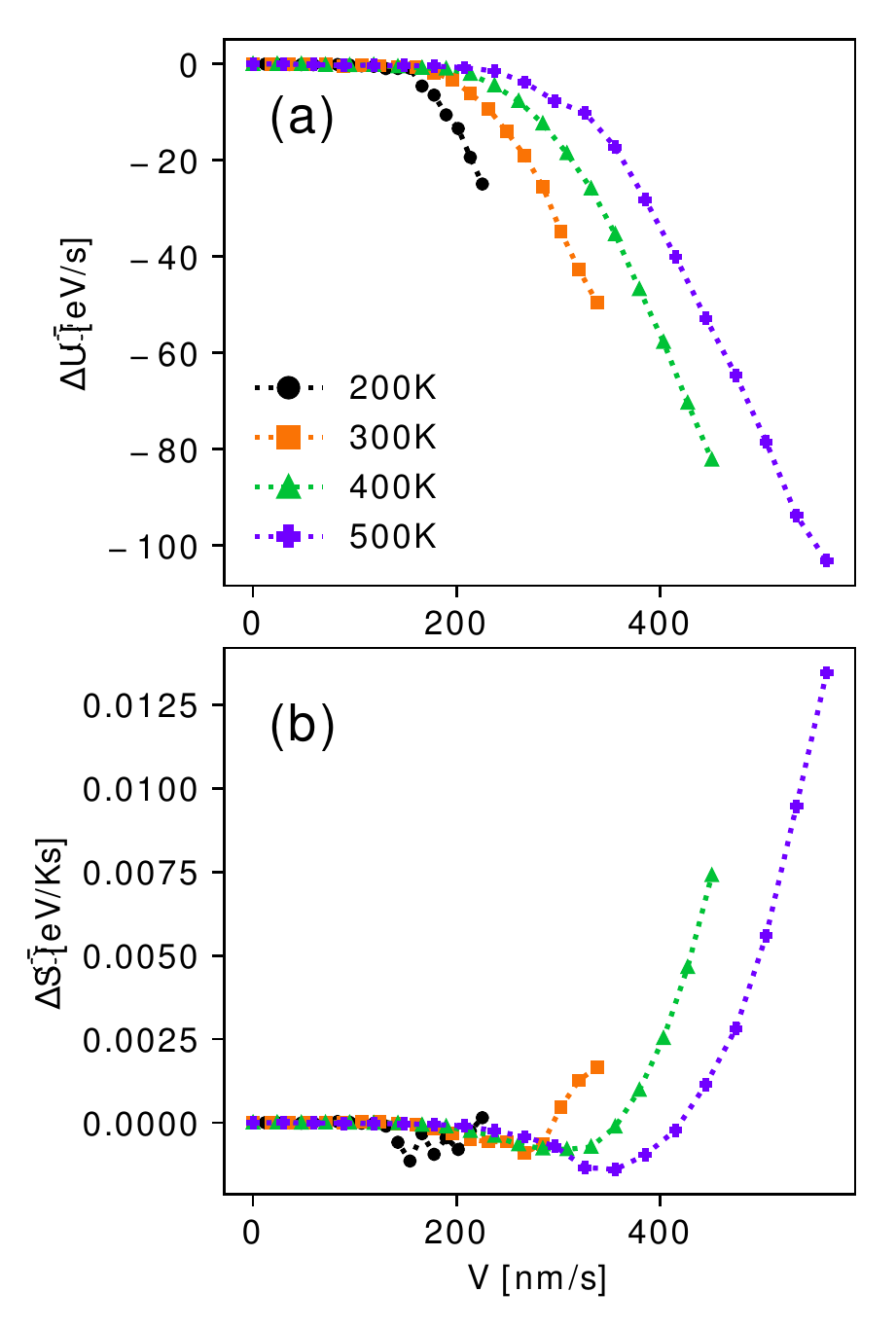}
  \caption{(a) Average internal energy change, (b) average entropy change, during stick-slip events, calculated for temperatures $200K-500K$, and velocities $0.1nm/s \leq V \leq V_{max}$ \cite{SI}.  The rest of model parameters are fixed at the default values.}
\label{fig:error}
\end{figure}

FIG. \ref{fig:error}(a) shows the internal energy change $\Delta\bar U$ computed as a difference between the average work per stick-slip event $\bar W = (1/\Delta t) \int_t^{t+\Delta t} \delta W/\delta t\,dt$ and the produced heat $\bar Q_p$. Although small for small velocities, $\Delta\bar U$ is non-zero in the entire velocity range. Thus the work performed on the system during the stick-slip event contributes both to the internal energy change and heat, and the thermodynamic formalism developed in previous sections is required for evaluating unambiguosuly their relative contributions. Similarly, in FIG. \ref{fig:error}(b) the average entropy change per stick-slip event $\Delta\bar S$, determined as a difference between $\bar Q_p/T$ and $\bar Q_e/T$, is non-zero and consequently the heat produced and transferred during the stick-slip event are not the same. 

The situation is simpler for a specific type of a cyclic process driven by periodically reversing velocity direction of the tip, when the friction losses can be determined directly from the area of the friction force loops. This can be shown by integrating Eqs. (\ref{eq:first_law}) and (\ref{eq:relation_three_entropies}) after a sufficient number of cyclic reversals of the velocity direction, when the system is expected to recover the same state after a velocity cycle and the cumulative changes of the state variables become zero, i.e. $\Delta\bar U|_\mathrm{cycle}=0$ and $\Delta\bar S|_\mathrm{cycle}=0$, which leads to relations $\bar W|_\mathrm{cycle} = \bar Q_e|_\mathrm{cycle}$ and $\bar Q_e|_\mathrm{cycle} = \bar Q_p|_\mathrm{cycle}$.

Finally, we comment on the range of validity of the Master equation formalism employed in this work. In the limit of high-energy barriers and over-damped dynamics, this approach is equivalent to Langevin dynamics, as has been shown earlier by a direct comparison between the solutions of the corresponding Fokker-Planck and Master equations \cite{evstigneev2015stochastic, evstigneev2004rate}. This work also demonstrated that the attempt frequency $f_0$, such as in Eq. (\ref{eq:transition-rates}), can be related to the effective mass, damping constant and sliding velocity of the cantilever \cite{comment}. The approach becomes problematic for fast dragging velocities when thermal fluctuations become irrelevant, which leads to deterministic dynamics. The sliding velocity threshold for this to occur can be estimated as discussed in \cite{SI}, confirming that the velocity range considered in FIG. \ref{fig:error} is within this threshold and is easily achievable by standard AFM equipment.



\section{\label{sec:discussion}Conclusion}

\noindent The developed formalism based on the Prandtl-Tomlinson model and the transition state theory incorporating the effects of thermal fluctuations allows to evaluate in full consistency the various relevant thermodynamic variables, including the internal energy, work, heat produced or transferred to the environment during the nanoscale friction process. We show that unless the friction losses are determined from the area of the friction force loops determined by cyclic FFM experiments using periodically reversed dragging velocity direction, explicit thermodynamic formulation is necessary for determining the friction losses during stick-slip events even in the steady state. 

The developed approach can be used for optimising the friction losses during arbitrary friction process, by determining the system parameters leading to specified levels of the produced heat. The approach can potentially be further generalised to increase the level of complexity, such as for Frenkel-Kontrova type models, or models based on solving the Langevin dynamics \cite{wang_energy_2015,pellegrini_thermally_2019}, which require evaluating the underlying Fokker-Planck equation. It is also naturally possible to extend the model to studying the 2D systems, if the underlying energy barriers and thermally activated transition paths can be identified by practical means. Also, different shapes of substrate potentials, such as determined directly by ab-initio methods can be incorporated into the analysis. Thus the present approach opens new directions for exploring the thermodynamics behaviour in a broad range of practically relevant materials.


\section*{Acknowledgments}
This work is supported by H2020 MSCA ITN project Solution No. 721642.


\bibliography{references}

\begin{thebibliography}{35}
\bibitem [{com()}]{comment}%
  \BibitemOpen
  \href@noop {} {}\bibinfo {note} {To illustrate that our calculations lead to
  realistic values, consider a slip event dynamics assuming a typical damping
  coefficient of $\eta\sim 10^{-9}$ kg/s and slip velocity $v\sim 1$ cm/s,
  which leads to an estimate of the kinetic energy dissipated during the event
  to be of the order of $\eta v^2\Delta t\sim 0.015$ eV, where we took $\Delta
  t= 2.5\times 10^{-8}$ s consistent with lattice spacing $0.25$ nm and slip
  velocity 1 cm/s. This estimate is within the order of magnitude consistent
  with our simulations, where for example in Fig. 5 the heat dissipated during
  one slip is of the similar order of 80 eV/s$\times$ 0.001 s $\sim$ 0.08
  eV.}\BibitemShut {Stop}%
\end{thebibliography}

\end{document}